\documentclass[prd,preprint,aps,showpacs]{revtex4}

\newcommand{\be}{\begin{equation}}
\newcommand{\ee}{\end{equation}}
\newcommand{\bn}{\begin{eqnarray}}
\newcommand{\en}{\end{eqnarray}}

\begin{document} 

\title{An analytical description for the cosmological constant}
\author{Everton M. C. Abreu and Carlos M. L. dos Reis}
\affiliation{Departamento de F\'{\i}sica, Universidade Federal Rural do Rio de Janeiro\\
BR 465-07, 23851-180, Serop\'edica, Rio de Janeiro, Brazil.\\
{\sf E-mail: evertonabreu@ufrrj.br}\\
\today}


\begin{abstract}
From the principles of quantum cosmologies we can justify the reason for an inverse-square law for the cosmological constant with no conflict with observations.  Although this general expression for $\Lambda$ is well known in the literature, in this work we  introduce some analytical solutions for the scale factor different from the literature.  The underlying motivation to carry out these solutions is the fact that the time variation of $\Lambda$ can lead to a creation of matter and/or radiation such as to help in the investigation of the age of the Universe.  The knowledge of the scale factor behavior might shed some light on these questions since the entire evolution of a homogeneous isotropic Universe is contained in the scale factor.
\end{abstract}

\maketitle



The research about the cosmological scale factor has attracted intense attention during the last years as well as the investigation about the cosmological constant problem.  As the vacuum has a non-trivial role in the early Universe, a $\Lambda$-term in the Einstein field equations is generated.  This $\Lambda$-term leads to the inflationary phase \cite{weimberg}.  Based on the inflationary cosmology we can say that during an early exponential phase, the vacuum energy was a large cosmological constant.  However, the current observed small value of the cosmological constant make us to assume that the cosmological constant $\Lambda$, representing  the energy density of the vacuum, is variable dynamic degree of freedom which being initially very large went down to its small present value in an expanding Universe \cite{av}.
The last one can be measured through the discrepancies between the infinitesimal value that the cosmological constant has for the present universe (it is very small in Planck units) and the values expected by the Standard Model \cite{berman}.

There are hundreds of works that show examples of phenomenological $\Lambda$-decay laws.  To sum up, the studies comprise a $\Lambda$ depending on temperature, time, Hubble parameter and scale factor \cite{oc}.  A dynamically decaying cosmological constant with cosmic expansion has been considered by several authors \cite{cw,ot,fafm,gasperini,clw,berman,kwe,abdel,beesham}.

The importance of the subject resides in the fact that any nonzero value of $\Lambda$ introduces a length scale and a time scale into the theory of general relativity \cite{bousso}.  The cosmological constant perturbs spacetime dynamics, although the general relativity works on scales much larger than the Planck scales.  However, vacuum quantum fluctuations of the Standard Model rekindle the theory of the cosmological constant problem.

Albeit there are many references about a scale factor inverse-square law for the cosmological constant there is a lack of information about the function of the scale factor that obey this requirement.  In reference \cite{abdel} there are some analytical solutions, however, those were obtained for a perfect-fluid matter energy-momentum tensor.  In this work we will use only the requirement of the energy-momentum conservation.

In this work we introduce some new analytical solutions for the scale factor different from the published and unpublished \cite{pk}  literature.  We find solutions that have different curvatures, $k=0,-1,+1$ as parameters, i.e., for flat, open and closed universes respectively.  The interest in a flat $(k=0)$ cosmological constant model has its motivation rooted in the fact that a $\Lambda$-term helps to connect inflation with observations.  Also that, with $\Lambda$, it is possible to obtain, for flat Universes, a theoretical age in the observed range, even for a high value of the Hubble parameter \cite{sw,cpt}.

To begin with some known calculations, the Einstein field equations are
\be
\label{1}
G_{\mu\nu}\,+\,\Lambda\,g_{\mu\nu}\,=\,8\,\pi\,G\,T_{\mu\nu}\,\,,
\ee
where $\Lambda$ has often been treated as a constant of nature and where $G_{\mu\nu}$ is the Einstein tensor given by 
\be \label{2}
G_{\mu\nu}\,=\,R_{\mu\nu}\,-\,{1\over 2}\,R\,g_{\mu\nu}
\ee
and $T_{\mu\nu}$ is the energy-momentum tensor.  We are using here relativistic units (i.e., $c=1$).

Taking the covariant divergence of (\ref{1}) and using Bianchi identities to guarantee the vanishing of the covariant divergence of the Einstein tensor, it follows (assuming the energy-momentum conservation law) that the covariant divergence of $\Lambda\,g_{\mu\nu}$ must vanish also, therefore $\Lambda=$const. constituting a geometrical interpretation of $\Lambda$.

As usual in recent works, we can move the cosmological term to the right-term in order to construct the Einstein equation with the form
\be
\label{3}
G_{\mu\nu}\,=\,8\,\pi\,G\,\tilde{T}_{\mu\nu} 
\ee
where
\be
\label{4}
\tilde{T}_{\mu\nu}\,\equiv\,T_{\mu\nu}\,-\,\frac{\Lambda}{8\pi G}g_{\mu\nu}
\ee
interpreting the cosmological term as part of the matter content of the universe, rather than a geometrical term.   With this effective momentum-tensor $\tilde{T}_{\mu\nu}$ satisfying the energy conservation as
\be
\label{5}
\nabla^{\nu}\,\tilde{T}_{\mu\nu}\,=\,0
\ee
there are no reasons in order to refuse a varying cosmological terms \cite{oc}.  
At this point we have to observe that since we use $c=1$, we will not consider here cosmological models with a time varying speed of light \cite{am}.

At this moment we have to make some initial assumptions such as a homogeneous and isotropic Universe.  In this case we will use the Robertson-Walker line element.  Let us consider also a perfect-fluid-like ordinary matter with pressure $p$ and energy density $\rho$ \cite{oc}.  The incorporation of the cosmological term in the definition written in (\ref{4}) implies that the effective energy-momentum tensor has the perfect fluid form, with an effective pressure given by $\tilde{p}\equiv p-\Lambda/8\pi G$ and an energy density as $\tilde{\rho}\equiv \rho+\Lambda/8\pi G$ \cite{oc}.

Using the field equations (\ref{3}) and (\ref{4}) and the law of energy-momentum conservation Eq. (\ref{5}) we can write
\be
\label{6}
\dot{a}^2\,=\,\frac{8\pi G}{3}\rho\,a^2\,+\,{\Lambda \over 3}a^2\,-\,k
\ee
and
\be \label{7}
\frac{d}{da}\left[\left(\rho+\frac{\Lambda}{8\pi G}\right)a^3\right]\,=\,-3\left(p\,-\,\frac{\Lambda}{8\pi G}\right)a^2\,\,.
\ee
For the equation of state we can write that 
\be \label{8}
p\,=\,(\gamma\,-\,1)\,\rho\,\,,
\ee
with $\gamma=$ constant.   Note from (\ref{8}) that for $\gamma=1$ we have matter-dominated epoch ($p=0$) and for $\gamma={4\over 3}$ we have radiation-dominated epoch.

Substituting (\ref{8}) in (\ref{7}) we have that
\be \label{9}
\frac{d}{da}(\rho a^{3\gamma})\,=\,-\frac{a^{3\gamma}}{8\pi G}\,\frac{d\Lambda}{da}\,\,,
\ee
and notice that when $\Lambda$=const. we have well known results \cite{oc}.
Now, using (\ref{6}) and (\ref{9}) we obtain that 
\be \label{10}
\ddot{a}\,=\,\frac{8\pi G}{3}\,\left(1-{{3\gamma}\over2}\right)\rho a \,+\,{\Lambda\over 3}a\,\,.
\ee

Using (\ref{6}) to eliminate $\rho$ in (\ref{10}) (or vice-versa) we can write the equation
\be \label{11}
{\ddot{a}\over a}\,=\,\left({\dot{a}}^2\,+\,k \right)\left(1\,-\,{3\gamma\over 2}\right){1\over a^2}\,+\,{\gamma\over 2}\Lambda\,\,,
\ee
which is a differential equation for the scale factor and has a cosmological term.  We clearly see that this equation is independent of the fact that $\Lambda$ is constant or not \cite{oc}.

The objective of this work is to solve equation (\ref{11}) for the scale factor.  But to accomplish this we have to choose a relevant expression for the cosmological term.

There is a whole literature that approaches this issue (see \cite{oc} and the references therein).  Here we choose a specific one (well known in the literature) but, as we said above, we will vary the curvature also.

We can summarize the forms of the cosmological constant depending on the scale factor in a general expression as \cite{matyjasek}
\be 
\label{11.1}
\Lambda(t)\,=\,\alpha\,a^{-m}\,+\,\beta\left({\dot{a}\over a}\right)^2\,\,,
\ee
where $\alpha$, $\beta$ and $m$ are constants.  Ozer and Taha \cite{ot} and Chen and Wu \cite{cw} concluded, although in different contexts, that the time dependence of $\Lambda$ should have $\beta=0$ and $m=2$ \cite{matyjasek}.  The interested reader can  see the different variations concerning these parameters in \cite{oc}.

Hence, let us choose the cosmological term as 
\be \label{12}
\Lambda\,=\,{\cal B}\,a^{-2}\,\,,
\ee
where ${\cal B}$ (we used here the same parameter used in \cite{oc}) is a pure number of order $1$ which should be calculable in a model for the time variation of $\Lambda$.   We can consider that the parameter ${\cal B}$ is a new cosmological parameter to be computed from observations, substituting $\Lambda$.  If ${\cal B}=0$ we have, obviously, an ansatz of $\Lambda=0$.  With a time variation like the exposed in (\ref{13}) the values of $\Lambda$ in the early Universe could be much bigger than the present value of $\Lambda$.  However, even that order of small magnitude it was arguably large enough to drive various symmetry breakings that occurred in the early Universe.  With a $\Lambda=0$ ansatz we cannot have such scenario.  Otherwise, if ${\cal B}\not=0$, the equation (\ref{13}) can be identified as a ``medium" time variation \cite{cw}.  For a positive ${\cal B}$ our Universe is flat.  In fact, considerations based on the second law of thermodynamics lead to the restriction ${\cal B}\geq0$ \cite{ot}.  Also, if ${\cal B}\geq k$ the Universe will expand forever; if ${\cal B}<k$ the Universe will collapse in the future.  For $\gamma>1$ in equation (\ref{11}), even a closed Universe will expand forever and the $k=0$ case is no longer critical for the collapse of our Universe \cite{ot}.
It can be shown that if our Universe is flat, observationally ${\cal B}$ should be positive \cite{cw}.   
The equation (\ref{12}) was also studied in \cite{oc,ln,ot,abdel,av2,cw,cops,mp}.  
The equation (\ref{12}) can be obtained theoretically from some assumptions from quantum cosmology.  It is not in conflict with experimental observations and with the inflationary scenario.  From quantum cosmology arguments it is more convenient to use the scale factor instead of the age $t$ of the Universe \cite{ot}.

In a general form of equation (\ref{12}) as a function of the scale factor $a^{-n}$, like the first term in equation (\ref{11.1}), the value $n=2$ has dimensional reasons.  With $n=2$, both $\hbar$ and $G$ disappear and we have, with $c=1$ only, the equation (\ref{12}) \cite{cw}.  The presence of $\hbar$ on the Einstein equations would be rather disturbing.

We will not make inflation considerations for the time being.  Equation (\ref{12}) can be obtained theoretically based on general assumptions on quantum gravity.  It can be considered also that this formulation has no conflicts with current data and can present some explanation about the inflationary scenario \cite{cw}.  In other words we can say that this model is singular and preserves the standard picture of the early Universe.  Notice that the condition $d\Lambda/da < 0$ requires that $B<0$ independently of the curvature index $k$ and 
hence $\Lambda>0$ for all $t\geq 0$ \cite{av3}.

The dependence on $a$ used in (\ref{12}) was first introduced by Gasperini \cite{gasperini} in a thermal approach.  A motivation for this kind of dependence is that it also appear in some string dominated cosmologies \cite{string}.   We can also obtain an Eq. (\ref{12}) form based on semiclassical Lorentzian analysis of quantum tunneling \cite{strominger}.  Although this kind of dependence is only phenomenological and does not come from particle physics first principles \cite{sw} it can be considered as a laboratory in order to investigate the models it generalizes.  As mathematically simple, we can obtain some analytical solutions.

\bigskip

After this brief historical review, let us now substitute Eq. (\ref{12}) in Eq. (\ref{11}) to have that
\be \label{13}
{\ddot{a}\over a}\,=\,\left(1\,-\,{3\gamma\over 2}\right)\,(\dot{a}^2\,+\,k)\,{1\over a^2}\,+\,{\gamma\over 2a^2}\,{\cal B}\,\,,
\ee
which can be written as in \cite{oc} as
\be \label{14}
a\,\ddot{a}\,+\,\Lambda_1\,\dot{a}^2\,+\,\Lambda_2\,=\,0\,\,,
\ee
where
\bn \label{15}
\Lambda_1&=&-\,\left(1\,-\,{3\gamma\over 2}\right)\,\,, \nonumber \\
\Lambda_2&=&-\,\left(1\,-\,{3\gamma\over 2}\right)\,k\,-\,{\gamma\over 2}\,B\,\,,
\en
and $\gamma$, $k$ and ${\cal B}$ constitute a group of free parameters which have the respective physical considerations mentioned above.  In our analysis, the parameter $k$, as said above, will have the values $0$, $-1$ and $+1$.  Just to remember, if we observe (\ref{12}) we can see clearly that if ${\cal B}=0$, then the expression covers the $\Lambda=0$ case.  If ${\cal B}\neq 0$ \cite{cw} we have two options, ${\cal B}\geq k$ where the Universe will expand forever or ${\cal B}<k$ and then the Universe will collapse in the future.   If ${\cal B}>1$, even a closed Universe (with $k=1$) will expand forever and the $k=0$ case is no longer critical for the collapse of our Universe \cite{cw}.

For the case $\Lambda_1=-1/2$ and substituting in (\ref{14}) we have that
\be \label{16}
a\,{d^2a \over dt^2}\,-\,{1\over 2}\left({da\over dt}\right)^2\,+\,\Lambda_2=0\,\,.
\ee

After some algebraic work we can reduce the solution to a simple form as
\be \label{18}
a(t)\,=\,-\,2\,{\Lambda_2\over c_1}\,+\,{c_1\over 4}\,(t\,+\,c_2)^2\,\,,
\ee
where $c_1$ and $c_2$ are constants to be determined. Observe that we have written the equation above in order to maintain the 
parameter $\Lambda_2$ in evidence for the following analysis below.

For this case of $\Lambda_1=-1/2$, from equations (\ref{15}) we have that $\gamma=1/3$ and consequently that $\Lambda_2$ has a dependence on $k$ and ${\cal B}$ that can be written as 
\be \label{18.1}
\Lambda_2=-{1\over 2}\,k\,-\,{1\over 6}\,{\cal B}\,\,,
\ee
and we have that for the values of $k=0,+1,-1$,
\bn \label{18.2}
\Lambda_2^{k=0}&=&-{1\over 6}\,{\cal B}\,\,, \nonumber \\
\Lambda_2^{k=-1}&=&{1\over 2}\,-\,{1\over 6}\,{\cal B}\,\,, \nonumber \\
\Lambda_2^{k=1}&=&-\,{1\over 2}\,-\,{1\over 6}\,{\cal B} \,\,,
\en
and the analysis about the expansion or collapse of the Universe can be accomplished as above.

Since the value ${\cal B}=0$ can be obviously discarded, we can stress that the general behavior of the solution (\ref{18}) does not  change.  As a final observation we can say for $\Lambda_1=-1/2$ the equation of state can be written as $p=-2/3$, where we have an inflation scenario.

For $\Lambda_1=-1$ we have two general solutions for Eq. (\ref{13}),
\bn \label{19}
a_{1}(t)\,=\,{1\over 4c_1} \left[e^{c_1(t+c_2)}\,-4\,\Lambda_2\,e^{-c_1(t+c_2)}\right]\,\,,  \nonumber \\
a_{2}(t)\,=\,{1\over 4c_1} \left[e^{-c_1(t+c_2)}\,-4\,\Lambda_2\,e^{c_1(t+c_2)}\right]\,\,,
\en
where the difference is the signs of the exponentials.

For $\Lambda_1=-1$ we have a little different behavior.   From (\ref{15}) we have that $\gamma=0$ and consequently that $\Lambda_2=-k$, so we can write that $\Lambda_2^{k=0}=0$, $\Lambda_2^{k=1}=-1$ and $\Lambda_2^{k=-1}=1$.   For $\Lambda_2=0$ the solutions (\ref{19}) have different behaviors altogether, namely,
\bn \label{19.1}
a_{1}(t)&=&{1\over{c_1}}e^{c_1 (t+c_2)}\,\,, \nonumber \\
a_{2}(t)&=&{1\over{c_1}}e^{-c_1 (t+c_2)}\,\,,
\en
and the equation of state now is $p=-\rho$, again an inflation scenario.

For $\Lambda_2=\pm 1$ we have that, 
\bn \label{19.2}
a_{1}^{k=\pm 1}(t)&=&{1\over 4c_1} \left[e^{c_1 (t+c_2)}\,\mp\,4\,e^{-c_1(t+c_2)}\right]\,\,, \nonumber \\
a_{2}^{k=\pm 1}(t)&=&{1\over 4c_1} \left[e^{-c_1 (t+c_2)}\,\mp\,4\,e^{c_1(t+c_2)}\right]\,\,,
\en
and we can see again completely different formulations for the scale factor in an inflation scenario.

For $\Lambda_1=1$ we have the following solution,
\be \label{20}
a(t)\,=\,\pm\sqrt{{c_1\over \Lambda_2}\,-\,\Lambda_2(t\,+\,c_2)^2}\,\,,
\ee
and from equations (\ref{15}) we have that $\gamma=4/3$ and 
\be \label{20.1}
\Lambda_2\,=\,k\,-\,{2\over 3}{\cal B}\,\,,
\ee
and the general behavior of the solutions (\ref{20}) do not change since we have the condition ${\cal B}\not= 0$.  However, we have to notice that the square root demands attention in order to not obtain imaginary scale factors for $k=0,\pm 1$.  But this is a trivial task to substitute in (\ref{20}) the values of $\Lambda_2$ obtained in (\ref{20.1}) with different $k$.  We will not have any new physical feature in this specific mathematical analysis.
The equation of state is $p={1\over 3}\,\rho$, a radiation-dominated epoch as said above.


For other small values of $\Lambda_1$ different from the ones analyzed above we do not have analytical solutions for the 
equation (\ref{14}).  For some values of $\Lambda_1$ we have singularities during the calculations that can not be circumvented, or the solutions can only be computed through numerical analysis.  And this is out of the scope of this paper.


\bigskip

As some final considerations we can say that in this paper we have computed some analytical solutions for the scale factor since the effective cosmological constant $\Lambda$ is time varying according to an inverse square law in the scale factor.  This inverse square law can be justified based on the principles of quantum cosmology.  Since such time variation of $\Lambda$ leads to creation of matter and/or radiation, we believe that the formulation of a scale factor analysis is of current interest both theoretical and experimental.  Another motivation can be found in the strict relation between the cosmological constant problem and the problem of the age of the Universe treated in the light of quantum cosmology.

There is a great literature about different functions for the cosmological constant.  Specifically concerning a function like in the equation (\ref{12}) the existing published papers do not bring any analytical expressions for the scale factor itself as a building term in $\Lambda$.  And the results found here, including the singularities, are in general different from the unpublished one \cite{pk}.

As we are interested in a general analytical behavior of $\Lambda$, numerical solutions of (\ref{14}) are out of the scope of this work.  Therefore we found three analytical solutions.   Other alternatives varying the parameter $\Lambda_1$ in (\ref{14}) brought singularities which demands numerical efforts.

Our main motivation is to show simply that there is another possible and different approach concerning the cosmological constant in order to obtain some progress in some questions about the Universe.  Although the model proposed by Chen and Wu in equation (\ref{12}) are compatible with the astrophysical data, it alters the predictions of the standard model for the matter-dominated epoch.  

As a future perspective it can be investigated, with the solutions obtained here, some aspects of inflation like the current value of the deceleration parameter and the parameter $\Omega_0$, namely, to obtain its value and test if they are compatible with the standard model demands.  Note that the knowledge of the form of the scale factor, as well known, can describe if an Universe is an expanding one or not through the line element.

An investigation about the age of the Universe in different scenarios as well as the equations of state, with the results obtained in this paper, is a work in progress.

\bigskip
\bigskip

The authors would like to thank Funda\c{c}\~ao de Amparo \`a Pesquisa do Estado do Rio de Janeiro (FAPERJ) and Conselho Nacional de Desenvolvimento Cient\' ifico e Tecnol\'ogico (CNPq) (Brazilian agencies) for financial support.

\end{document}